\documentclass[11pt,letterpaper]{article}
\usepackage[latin1]{inputenc}
\usepackage{subfigure}
\usepackage{lineno}
\usepackage{geometry}  
\usepackage{amsmath}
\usepackage{amsfonts}
\usepackage{amssymb}
\usepackage{graphicx}
\usepackage{authblk}
\usepackage{braket}
\usepackage{scalerel}
\usepackage{color}
\definecolor{ascolor}{rgb}{1,0,1}

\definecolor{kkcolor}{rgb}{1,0,0}

\DeclareRobustCommand\asout{\bgroup\markoverwith{\color{ascolor}{\rule[0.4ex]{2pt}{0.8pt}}}\ULon}
\newcommand{\asb}{\bar{\alpha}_s}
\newcommand{\bqt}{{\bf q}_{\scaleto{T}{4pt}}}
\newcommand{\bkt}{{\bf k}_{\scaleto{T}{4pt}}}
\newcommand{\bkto}{{\bf k}_{\scaleto{T0}{4pt}}}
\newcommand{\qt}{{q}_{\scaleto{T}{4pt}}}
\newcommand{\kt}{{k}_{\scaleto{T}{4pt}}}
\newcommand{\kto}{{k}_{\scaleto{T0}{4pt}}}
\newcommand{\Qo}{{Q}_{\scaleto{0}{4pt}}}
\newcommand{\cF}{{\cal F}}

\title{\bf On the different forms of the kinematical constraint in BFKL}

\author[1]{Michal Deak}
\author[2]{Krzysztof Kutak}
\author[1]{Wanchen Li}
\author[1]{ Anna M. Sta\'sto}

\affil[1]{\small \it Department of Physics, Penn State University, University Park, PA 16802, USA}
\affil[2]{\small \it H. Niewodniczanski Institute of Nuclear Physics, Polish Academy of Sciences, Krakow, Poland}

\begin{document}
\maketitle
\begin{abstract}
We perform a detailed analysis of the different forms of the kinematical constraint imposed on the low $x$ evolution that appear in the literature.
We find that all of them generate the same leading anti-collinear poles in Mellin space which agree with BFKL up to NLL order and up to NNLL in $N=4$ sYM.
The coefficients of subleading poles vanish up to NNLL order for all constraints and we prove that this property should be satisfied to all orders.  We then demonstrate  that the kinematical constraints differ at further subleading orders of poles. We quantify the differences between the different forms of the constraints by performing numerical analysis both in Mellin space and in momentum space.  It can be shown that in all three cases BFKL equation can be recast into the differential form, with the argument of the longitudinal variable shifted by the combination of the transverse coordinates. 

\end{abstract}
\section{Introduction}

The sharp increase of the structure function $F_2$ in  Deep Inelastic Scattering process   with decreasing value of Bjorken variable $x$   was one of the biggest discoveries at HERA collider \cite{Abt:1993cb,Derrick:1993fta}.  This rise of $F_2$ is  being driven in QCD by the increase in the  gluon density due to dominance of gluon splitting when the parton's longitudinal momentum fraction is small. Understanding  the behavior of the parton densities at small $x$ is not only crucial for the DIS processes but also for the high energy hadronic collisions in accelerators and  in the cosmic ray interactions. The LHC opened up the kinematic regime which is particularly sensitive to small $x$ values and thus the region where the  gluon density dominates. The increase of the gluon density towards small $x$  translates 
then into the rise of the hadronic cross sections for variety of observables  with the increasing center-of-mass energy. Within the  framework of collinear factorization  \cite{Collins:1989gx} the evolution of the gluon density is governed by the DGLAP \cite{Gribov:1972ri,Altarelli:1977zs,Dokshitzer:1977sg} evolution equations which lead to the scaling violations.  Although  the DIS data are well described by the fits based on the DGLAP evolution, there are hints of novel physics phenomena that can be present at very small values of $x$ \cite{Ball:2017otu}.  The alternative approach to the description of the parton evolution stems from the analysis of the Regge limit in QCD, where the energy of the interaction is assumed to be  large $s\rightarrow \infty$, or in the case of the DIS process, when $x\rightarrow 0$, while the scale $Q^2$ is perturbative but fixed. In this limit the cross section is dominated by the exchange of the so-called hard Pomeron which is the solution to the famous BFKL evolution equation \cite{Lipatov:1985uk,Kuraev:1977fs,Balitsky:1978ic}, see \cite{Lipatov:1996ts} for a review. The two notable differences between DGLAP and BFKL evolution equations are that in the latter the solution exhibits the power like behaviour, $x^{-\lambda}$ at small values of $x$ and the transverse momenta in  the gluon cascade are not ordered. The solution thus depends on the transverse momenta of the exchanged reggeized gluons. BFKL equation is known up to NLL order in QCD \cite{Fadin:1998py,Ciafaloni:1998gs} and up to NNLL order in $N=4$ sYM  theory \cite{Gromov:2015vua,Velizhanin:2015xsa,caron2018high}. A known issue with the  BFKL  at the NLL order is that the corrections are large compared to the LL result, which renders the result unstable. Thus it was early realized that higher order corrections need to be resummed.
In the pioneering works \cite{Ciafaloni:1987ur, Andersson:1995ju,Kwiecinski:1996td} it was pointed out that there are kinematical constraints which need to be included in the evolution. Formally of higher order, when viewed from the perspective of the Regge limit, nevertheless such corrections are numerically very large as was established in \cite{Kwiecinski:1996td}. Kinematical constraint was later included in the formalism that combined DGLAP and BFKL equations, and it was demonstrated that very good description of the HERA data was obtained \cite{Kwiecinski:1997ee}. Various elaborated resummation schemes for the small $x$ evolution were later developed  \cite{Salam:1998tj, Altarelli:1999vw,Altarelli:2000mh,Altarelli:2001ji,Altarelli:2003hk,Altarelli:2008aj,Ciafaloni:1999au,
	Ciafaloni:1999yw, Ciafaloni:2003rd,Ciafaloni:2003kd,Ciafaloni:2007gf,Vera:2005jt,Bonvini:2016wki,Thorne:2001nr,White:2006yh}.
More recently the collinear resummation was applied to the non-linear evolution equations in transverse coordinate space \cite{Motyka:2009gi, Beuf:2014uia, Iancu:2015vea,Hatta:2016ujq}.

Given the increased precision of the experimental data and progress in theoretical computations, especially the information about the higher orders in $N=4$ sYM theory, it is worth to revisit the assumptions in the resummed schemes. Even though the kinematical constraint is well motivated, different approximations, which are used are usually  done based on general grounds (i.e. for example low $z$ approximation) but without detailed quantitative analysis. The differences between different forms of constraints may be of subleading order but it could happen that they are quite large numerically. In  context of the BFKL equation, which utilizes the kinematical constraint as an ingredient, usually one uses a  particular form of this constraint, namely the one which limits the transverse momenta of the exchanged gluons. Such choice is also dictated by practical applications, since in this form the angular variable in the evolution equation is integrated over which reduces the number of variables and makes it easier to solve  the equation. On the other hand, there are different forms of the kinematical constraint that appear in the literature,
which put limit on the transverse momenta of the emitted gluons. Actually in the CCFM equation which receives contribution from both  large and low $x$ it is a natural choice  \cite{Hautmann:2014uua}. \footnote{All the forms of the kinematical constrains were implemented in numerical study of the CCFM equation \cite{Hautmann:2014uua}.}   

The main goal of this paper is to perform a systematic semi-analytic and numerical comparison of the small $x$ evolution equation with different forms of the kinematical constraint, which appear in the literature. 

In the analysis performed in this work we found that the constrains all generate the same structure in the Mellin space for the leading and first subleading  anti-collinear poles up to NLL order in $\ln 1/x$ in QCD  and in $N=4$ sYM and up to NNLL in $N=4$ sYM.  As such they are all consistent up to the highest known order of perturbation theory.   The calculations were performed both in the Mellin space and also by  performing the direct solution of the BFKL with kinematical constraint in the momentum space. To this aim we have developed new numerical method for the solution of the BFKL equation with kinematical constraint which is based on reformulating this equation in the differential form. It turns out that all the constraints can be recast as effective shift of the longitudinal momentum fraction in the argument of the unintegrated gluon distribution.  We see that, as expected the solutions are substantially reduced with respect to the LL case, but the differences between the different forms of the kinematical constraints are non-negligible and can reach up to 10-15 \% for the value of the intercept. This can have important effect for the phenomenology.

\section{Kinematical constraints}
\label{sec:constraints}
	
In this section we shall review the origin and different forms of the kinematical constraints that appear in the literature.
The kinematical constraint in the initial state cascade at low $x$ was first considered in works \cite{Ciafaloni:1987ur,Catani:1989sg, Andersson:1995ju}.

\begin{figure}
	\centerline{\includegraphics[width=5cm]{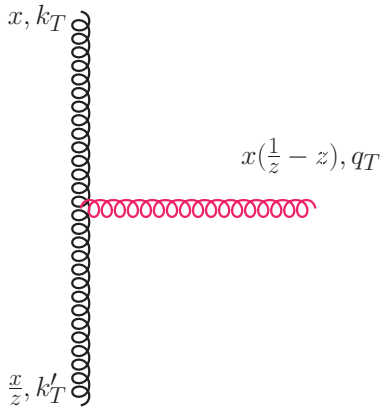}}
	\caption{Schematic diagram representing gluon emission in the BFKL chain. $x$ and $x/z$ are the longitudinal momentum fractions of the target's momentum carried by the exchanged gluon. $\bkt,\bkt'$ and $\bqt$ denote the two-dimensional transverse momenta of the exchanged gluons and the emitted gluon respectively.}
	\label{fig:diag1}
\end{figure}

We shall follow here closely the derivation presented   in \cite{Kwiecinski:1996td}, where the kinematical constraint was implemented  both in BFKL and the CCFM equations \cite{Ciafaloni:1987ur,Catani:1989sg,Catani:1989yc,Marchesini:1994wr,Jung:2010si}. The latter evolution equation  is based on the idea of coherence \cite{Dokshitzer:1987nm} that leads to the angular ordering of the emissions in the cascade.
The BFKL equation for the unintegrated gluon density in the leading logarithmic approximation \cite{Balitsky:1978ic,Kuraev:1977fs,Lipatov:1985uk} can be written as

\begin{multline}
\cF(x,\kt^2)=\cF^{(0)}(x,\kt^2)\\
+\asb\int_{x}^{1}\frac{dz}{z}\int\frac {d^2 \bqt}{\pi  \qt^2}\left[\cF\left(\frac{x}{z},
\left|\bkt+\bqt\right|^2\right)-\Theta(\kt^2-\qt^2)\cF\left(\frac{x}{z}, \kt^2\right)\right] \; ,
\label{eq:bfkleq}
\end{multline}
where function  $\cF(x,\kt^2)$ is the small $x$ unintegrated gluon density, $\bkt$ and $\bkt'=\bkt+\bqt$ are the transverse momenta of the exchanged gluons
and $\bqt$ is the transverse momentum of the gluon emitted. The longitudinal momentum fractions of the exchanged gluons are $x$ and ${x \over z}$ respectively.
The flow of the momenta in the BFKL cascade is illustrated in diagram Fig.~\ref{fig:diag1}. 
We will also use the notation $\kt^2\equiv\bkt^2$ for the squared transverse momenta for the rest of the paper. The rescaled coupling constant is defined as $\asb \equiv \frac{\alpha_s N_c}{\pi}$.
In the leading logarithmic approximation the 
integration over ${\bqt}$ in the Eq.~\eqref{eq:bfkleq},
 is not 
constrained by an upper limit thus violating the energy-momentum conservation. 
Thus for example in the context of the DIS process in principle there should be a limit on the integration over the transverse momentum
\begin{equation}
 Q^2/x  \sim W^2\, ,
\end{equation}
where $Q^2$ is the hard scale of the DIS process,  $x$ is the Bjorken variable and $W^2$ is the c.m.s energy squared of the photon-proton system in DIS. 

There is a stronger constraint arising however from the requirement that  in the low $x$ formalism 
the exchanged gluons have off-shellness dominated by the transverse components, i.e. one keeps terms that obey $|k^2| \simeq \kt^2$. The derivation of the kinematical constraint here follows \cite{Kwiecinski:1996td}.

The gluon four momentum is as usual decomposed into light cone and transverse components
\begin{equation}
k = (k^+,k^-,\bkt) \, ,
\end{equation}
where $k^{\pm}=k^0\pm k^3$.
Now the exchanged gluon virtuality in these variables is equal to
\begin{equation}
k^2 = k^+k^- - \kt^2 \;.
\end{equation}

The condition $|k^2|\simeq \kt^2$ translates approximately to
\begin{equation}\label{eq:ksqs0app}
\kt^2> k^+ k^- \, .
\end{equation}
The emitted gluon is on-shell $q^2=q^+ q^- -\qt^2=0$, and therefore we can express  this  fact as
\begin{equation}
q^-=\qt^2/q^+ \, .
\label{eq:onshellq}
\end{equation}
On the other hand in the multi-Regge kinematics there is a strong ordering of longitudinal momenta so
\begin{equation}
k^- = k^{'-}-q^{-} \simeq q^- \,.
\end{equation}
Using the condition \eqref{eq:onshellq} and inserting   it into~\eqref{eq:ksqs0app} one finally obtains
\begin{equation}\label{eq:kincon-fa}
\kt^2>k^+ \frac{\qt^2}{q^+} =\frac{z}{1-z} \qt^2\;,
\end{equation}
or as limit on $\qt^2$ integration
\begin{equation}\label{eq:kincon-f}
\qt^2<\frac{1-z}{z} \kt^2\;.
\end{equation}
Now there are several approximations that can be made to this constraint.
In the small $z$ limit ~\eqref{eq:kincon-f}  can be approximated to 
\begin{equation}
\qt^2 < \frac{\kt^2}{z}\, .
\label{eq:kincon-jk}
\end{equation}
This form of the approximation was used in \cite{Ciafaloni:1987ur}  and also studied  in the context of small $x$ approximation to the CCFM evolution in  \cite{Kwiecinski:1996td}.
The lower bound on $z$, i.e. $z>x$ results in the upper bound on 
$\qt^2<\kt^2/x$ providing local condition for energy-momentum 
conservation.

Finally,  (\ref{eq:kincon-f}) can be further rewritten as a condition on the transverse momentum of the exchanged gluon $\kt'$. For a given value of $\kt$ a high value of $\kt'$ means also high value of $\qt$. Rewriting it as
\begin{equation}
\kt^{\prime 2}-2\kt^{\prime}\cdot\kt+\kt^{2}<\frac{1-z}{z} \kt^2\, ,
\label{eq:kinconorig}
\end{equation}
and averaging over angle between $\kt$ and $\kt^{\prime}$ and taking large $\kt^{\prime}$ limit we get:
\begin{equation}
\kt^{\prime 2} <  \frac{\kt^2}{z} \,.
\label{eq:kinconorig}
\end{equation}
This form of the constraint was used for example in \cite{Andersson:1995ju} and in \cite{Kwiecinski:1997ee,Ciafaloni:2003ek,Ciafaloni:2003rd,GolecBiernat:2001if,Kutak:2003bd}. 
The nice feature of \eqref{eq:kinconorig} is the fact that the kernel with kinematical constraint has a Mellin representation which results in a simple shift of poles in the Mellin  space.
In the rest of the paper we shall analyze in detail all the forms of the constraints and quantify the differences between them both in Mellin space and through direct numerical solution ot the BFKL equation in momentum space.
	\section{Comparison of improved kernel with results in $N=4$ sYM}
	\label{sec:n4sym}
Since the works of \cite{Andersson:1995ju,Kwiecinski:1996td}  it is known that the kinematical constraint provides very important contribution  to the NLL order and beyond in BFKL. Even though formally of subleading order, numerically this correction is very large and leads to the strong reduction of the intercept, aligning the BFKL equation more with phenomenology, for selected early works which implement kinematical corrections in BFKL see for example \cite{Kwiecinski:1997ee,  Orr:1997im, Andersen:2003gs,Avsar:2005iz}.  At the NLL order the kinematical constraint, when viewed as a correction in the Mellin space  gives rise to the cubic poles in the Mellin conjugated variable to the transverse momentum in the BFKL eigenvalue. The $N=4$ sYM eigenvalue at LL is identical to the QCD case, and at NLL  the same leading cubic poles appear in both theories \cite{kotikov_dglap_2003}. Since the  constraint is originating from the kinematics specific to the Regge limit of the cascade, one could expect it to be universal in both theories. It is useful to explore whether or not  the kinematical constraint correctly reproduces the terms in the higher order NNLL known only in the supersymmetric case \cite{Gromov:2015vua,Velizhanin:2015xsa,caron2018high} and what are the differences between the different forms of the kinematical constraint discussed in the literature. 

In this section we shall perform a detailed comparison of leading and subleading poles in the Mellin space originating from the kinematical constraint with the results in $N=4$ sYM case up to NNLL.
First, we shall perform the comparison on  the example of the constraint \eqref{eq:kinconorig}.  Later on, in Sec.~\ref{sec:PropsEigenvs},  we shall check that the results are consistent for different forms of the kinematical constraints up to certain order of poles, and we shall identify this order. In that way we can quantify the level of differences between different forms of the constraints.

\subsection{NLL and NNLL of N=4 sYM and $\omega$ expansion}

Let us begin with recalling the definition of the BFKL eigenvalue and the Mellin transform.

The Mellin transformation into the  \(\omega\) space is defined as
\begin{equation}
\overline{\cal F}(\omega,k_T^2)= \int_{0}^{1}\frac{dz}{z}z^{\omega}\, \cF(z,\kt^2) \; ,
\end{equation}
and into the \(\gamma\) space as
\begin{equation}
\tilde{\cF}(\omega,\gamma)= \int_{0}^{\infty} d\kt^2 \left(\kt'^2\right)^{-\gamma} \cF(\omega,\kt^2) \; ,
\end{equation}
Applying it to the BFKL equation  we get the algebraic form of the BFKL equation
\begin{equation}
\tilde{\cF}(\omega,\gamma)=\tilde{\cF}^{(0)}(\omega,\gamma) + \frac{\asb}{\omega} \chi(\gamma,\omega) \tilde{\cF}(\omega,\gamma) \, ,
\end{equation}
where $\chi(\gamma,\omega)$ is the kernel transformed to the double Mellin space.
The $\omega$ dependence stems from the $z$ dependence of the kernel $K$ due to the kinematical constraint.
Of course the fixed order LL and NLL kernels do not have the $\omega$ dependence, it is only the resummed kernel, in this case the kernel with the  kinematical constraint that obtains this additional dependence.  As is well known, see for example \cite{Kwiecinski:1996td,Kwiecinski:1997ee,Salam:1998tj,Ciafaloni:2003ek,Ciafaloni:2003rd}, such dependence in turn generates the series in $\alpha_s$ and therefore resums the contributions from all orders in the strong coupling.  The result with the constraint \eqref{eq:kinconorig}
is well known \cite{Andersson:1995ju}
\begin{equation}
\chi(\gamma,\omega) =    2 \psi(1) - \psi(\gamma) - \psi(1-\gamma+\omega) \, .\label{eq:shifted_asym}
\end{equation}
where $\psi$ is a polygamma function. When $\omega=0$ the above result coincides with the LL BFKL eigenvalue in QCD and in N=4 sYM
\begin{equation}
\chi_{0}(\gamma) =    2 \psi(1) - \psi(\gamma) - \psi(1-\gamma) \, .\label{eq:llx}
\end{equation}
We shall investigate Mellin forms of the kernels with other constraints \eqref{eq:kincon-f} and \eqref{eq:kincon-jk} later on in Sec.\ref{sec:PropsEigenvs}.
In order to compare the results with NLL and NNLL calculations one needs to specify the correct scale choice.
The above result leads to an asymmetric kernel which is valid in the so-called asymmetric scale choice. That means it is valid when  considering the DIS process, in which the $x$ Bjorken is defined
as $x=Q^2/s$ with $Q^2$ the (minus) virtuality  of the photon. For the case of the different process, like for example Mueller-Navelet jets with comparable transverse momenta, the appropriate variable would  be $QQ_0/s$, where $Q \simeq Q_0$ are the scales of the order of transverse momenta of the jets.  The eigenvalue in this case would be different from \eqref{eq:shifted_asym} as it should correspond
to the symmetric choice of scales. Therefore one needs to perform the scale changing transformation which generates terms starting at an appropriate  order. Up to NLL order this was discussed in  \cite{Ciafaloni:1998gs, Fadin:1998py, Salam:1998tj}.
We shall recall in detail the scale changing transformation, and what terms it generates up to NNLL in the next subsection, here we shall directly start from the symmetric
counterpart of the eigenvalue \eqref{eq:shifted_asym} which has the following form
\begin{equation}
\chi(\gamma,\omega) =    2 \psi(1) - \psi(\gamma + \frac{\omega}{2}) - \psi(1-\gamma+\frac{\omega}{2}) \, .\label{eq:shifted_sym}
\end{equation}
The result for the NNLL eigenvalue in the case of the $N=4$ sYM was derived originally in \cite{Gromov:2015vua,Velizhanin:2015xsa} for the symmetric case. It was later on rederived in \cite{caron2018high} by exploiting the correspondence between the soft-gluon wide-angle radiation in jet physics and the BFKL physics. To have direct relation to these results we will focus below on the symmetric case.
The poles around  $\gamma=0$ of NLL \cite{kotikov_dglap_2003} and NNLL kernel in N=4 sYM are given by \cite{Gromov:2015vua,Velizhanin:2015xsa,caron2018high}
\begin{align}
&\chi_1^{sYM}=-\frac{1}{2\gamma^3}-1.79+{\cal O}(\gamma) \; ,\\
&\chi_2^{sYM}=\frac{1}{2\gamma^5}-\frac{\zeta(2)}{\gamma^3}-\frac{9\zeta(3)}{4\gamma^2}-\frac{29\zeta(4)}{8\gamma}+{\cal O}(1) \; .
\label{eq:polesn=4sym}
\end{align}
Since it is symmetric case the coefficients of the poles around $\gamma=0$ and $\gamma=1$ are identical. For the purpose of simplification therefore we only focus on the expansion around $\gamma=0$.
We can retrieve the leading and the vanishing subleading poles of the \(N=4\) sYM case by doing \(\omega\) expansion of the shifted eigenvalue \eqref{eq:shifted_sym}
\begin{equation}
\chi(\gamma,\omega)=2\psi(1)-\psi(\gamma+\frac{\omega}{2})-\psi(1-\gamma+\frac{\omega}{2})
=\chi_0+\chi^{(1)}\frac{\omega}{2}+\frac{1}{2!}\chi^{(2)}\left(\frac{\omega}{2}\right)^2+\dots \; ,
\label{omega expansion}
\end{equation}
where the \(\chi^{(i)}\) is the i-th derivative of \(\chi^{\omega}\) with respect to \(\omega\). The  term lowest in order in $\asb$ in the expansion is simply
$\chi_0$  which is the LL eigenvalue \eqref{eq:llx}. In order to retrieve the  term contributing to the NLL order
we use the solution to the  equation for the intercept in the lowest order of the  coupling, i.e. \(\omega_0=\asb \chi_0\), and substitute it  into \eqref{omega expansion} and keep terms up to first power in $\asb$. One obtains
\begin{equation}
\chi(\gamma)=\chi_0+\frac{1}{2}\asb\chi^{(1)}\chi_0 \; .
\end{equation}
This gives the contribution to the NLL order
\begin{equation}
\chi_1(\gamma)=\frac{1}{2}\chi^{(1)}\chi_0=\frac{1}{2}\left[\psi^{(1)}(\gamma)+\psi^{(1)}(1-\gamma)\right]\left[2\psi(1)-\psi(\gamma)-\psi(1-\gamma)\right]\, .
\end{equation}
Expanding around $\gamma=0$ one obtains the following pole structure
\begin{equation}
\chi_1(\gamma)=-\frac{1}{2\gamma^3}-\frac{\zeta(2)}{\gamma}+O(1)\, .
\label{eq:chi1omega}
\end{equation}

Similarly, we can derive the  term contributing at NNLL level from the \(\omega\) expansion to the second order by taking\footnote{The subscript on $\omega$ indicates the order of expansion in which we are interested in.} $\omega_1=\asb\chi_0+\asb^2 \chi_1$, substituting again into \eqref{omega expansion} and extracting terms proportional to $\asb^2$ which results in
\begin{equation}
\chi_2=\frac{1}{4}\left(\chi^{(1)}\right)^2\chi_0+\frac{1}{8}\chi^{(2)}\left(\chi_0\right)^2 \; .
\end{equation}
In this case the pole structure is
\begin{equation}
\chi_2=\frac{1}{2\gamma^5}+\frac{\zeta(2)}{\gamma^3}+\frac{2\zeta(3)}{\gamma^2}+O\left(\frac{1}{\gamma}\right) \; .
\label{eq:chi2omega}
\end{equation}
We observe that the leading poles $-1/2\gamma^3$ \eqref{eq:chi1omega} and $1/2\gamma^5$ \eqref{eq:chi2omega} coincide with the exact result at NLL  in $N=4$ SYM and in QCD  and NNLL order in $N=4$ SYM \eqref{eq:polesn=4sym}. This structure is consistent with the principle of maximal transcendentality (complexity) \cite{kotikov_dglap_2003} meaning that all special functions at the NNLO correction contain only sums of the terms $~1/\gamma^i$ (i= 3, 5).
What is also interesting is the absence of subleading poles, i.e. $1/\gamma^2$ and $1/\gamma^4$ at NLL and NNLL respectively  in the 
$\omega$ expansion of kinematical constraint which again coincides with  the exact results in $N=4$ sYM \eqref{eq:polesn=4sym}. We shall see later in this  Section  that one expects this pattern is expected to hold to all orders in $\asb$ from the kinematical constraint. Of course in the case of QCD  at NLL there are double poles $1/\gamma^2$, but they originate from the non-singular parts of the QCD DGLAP anomalous dimension and the running coupling \cite{Ciafaloni:1998gs,Fadin:1998py}.

\subsection{Scale changing transformation at NNLL}

In this section we shall recall the scale changing transformation and compute it to NNLL order. 
The scale changing transformations were discussed in \cite{Ciafaloni:1998gs}, see also \cite{Ciafaloni:2003ek,Ciafaloni:2003rd}.

Let us start with the general formula in the high energy factorization for the process with two scales $Q,Q_0$

\begin{equation}
\sigma=\int\frac{d\omega}{2\pi i}\int \frac{d^2 \bkt }{\kt^2}\frac{d^2 {\bkto}}{{\kto}^{\!\!\!2}}\left(\frac{s}{Q\Qo}\right)^\omega h^A(Q, \bkt)\,G_\omega(\bkt,\bkto)\,h^B(\Qo,{\bkto}) \, .
\label{eq:xsection}
\end{equation}

Here, $h^A$ and $h^B$ are impact factors, that depend on the process in question and $G_\omega(\bkt, \bkto)$ is the BFKL gluon Green's function.
The unintegrated gluon density introduced in the previous section can be interpreted ad the convolution of the Green's function with one of the impact factors, i.e. 
\begin{equation}
\bar{\cF}(\omega,\bkt) \; = \; \int \frac{d^2 {\bkto}}{{\kto}^{\!\!\!2}}\,G_\omega(\bkt,\bkto)\,h(\Qo,{\bkto}) \; .
\end{equation}

 In Eq.\eqref{eq:xsection} $Q$,$\Qo$ are hard scales for this process. For example in Mueller-Navelet process of production of two jets with comparable transverse momenta $Q\approx \Qo$. On the other hand in  DIS $Q\gg \Qo$. The equation for $G$ as a function of rapidity is
\begin{equation}
G(Y;\bkt,{\bkt}_0)=\int \frac{d\omega}{2\pi i}\,e^{\omega Y} \,G_\omega(\bkt,{\bkt}_0)\, ,
\end{equation}
where $Y=\ln(\nu/\kt\kto)$, $\nu$ is the energy available for the BFKL evolution in a given process.

Here we see that the choice of the energy scale is symmetric $\nu/\kt  \kto$, but it could also be asymmetric $\nu/\kt^2$ or $\nu/\kto^2$. This will imply that 
\begin{equation}
G_\omega\rightarrow G_\omega\left(\frac{k_{>}}{k_{<}}\right)\, ,
\end{equation}
where $k_>=\max(\kt,\kto)$ and $k_<=\min(\kt,\kto)$.
For example:
\begin{equation}
\int \frac{d\omega}{2\pi i}\left(\frac{\nu}{\kt \kto}\right)^\omega G_\omega(\bkt,\bkto)=\int \frac{d\omega}{2\pi i}\left(\frac{\nu}{\kt^2}\right)^\omega \left(\frac{\kt}{\kto}\right)^\omega G_\omega(\bkt,\bkto)\; .
\end{equation}
This means that depending on the scale choice the kernel will undergo a scale changing transformations as well.
\begin{eqnarray}
K_\omega^{s_0=\kt^2}(\bkt,\bkt0)&=K_\omega^{s_0=\kt\kto}(\bkt,\bkto)\left(\frac{\kt}{\kto}\right)^\omega\, ,\\
K_\omega^{s_0=\kto^2}(\bkt,\bkt0)&=K_\omega^{s_0=\kt\kto}(\bkt,\bkto)\left(\frac{\kto}{\kt}\right)^\omega\; .
\end{eqnarray}
In the Mellin space 
this will imply the shift of poles in $\gamma$  by $\omega$ for the  function $\chi^{\omega}(\gamma)$. 
This can be expressed as \cite{Ciafaloni:1998gs}
\begin{equation}
\chi^S(\gamma')=\chi^S\left(\gamma-\frac{\omega}{2}\right)=\chi^A(\gamma) \,.
\end{equation}

So for example in the case of kernel with kinematic constraint $k'^2<k^2/z$, we have in the asymmetric case
\begin{equation}
\chi^{\omega,A}(\gamma)=2\psi(1)-\psi(\gamma)-\psi(1-\gamma+\omega)\; ,
\label{eq:chi0a}
\end{equation}
and in the symmetric case,
\begin{equation}
\chi^{\omega,S}(\gamma)=2\psi(1)-\psi\left(\gamma+\frac{\omega}{2}\right)-\psi\left(1-\gamma+\frac{\omega}{2}\right)\;.
\label{eq:chi0s}
\end{equation}

Let us now recall how the scale changing works at NLL before proceeding to NNLL. We suppose that in  the following an additional dependence on $\omega$  might be present in the $\chi$ function. Going from symmetric case to an  asymmetric case we have at this order
\begin{equation}
\chi_0\left(\gamma-\frac{\omega}{2}\right)+\asb\chi_1\left(\gamma-\frac{\omega}{2}\right).
\label{SYM to AYSM}
\end{equation}
Expanding in $\omega$ and keeping terms
up to NLL we get
\begin{equation}
\chi_0(\gamma)-\frac{\omega}{2}\frac{\partial \chi_0}{\partial \gamma}+\asb \chi_1(\gamma)\, .
\end{equation}
The scale changing part is given by 
\begin{equation}
T^{\rm NLL}(\gamma) = -\frac{\omega}{2}\frac{\partial \chi_0}{\partial \gamma} \; , 
\end{equation}
and at  this order we need to take $\omega=\asb \chi_0(\gamma)$, therefore scale changing part is
\begin{equation}
T^{\rm NLL}(\gamma) =   -\frac{1}{2}\asb\chi_0(\gamma)\frac{\partial \chi_0}{\partial \gamma} \; ,
\end{equation}
see \cite{kotikov_dglap_2003,Fadin:1998py,Ciafaloni:1998gs}. In the following, we will focus only on the pole structure and  keep leading poles in $\gamma$ which gives for the scale changing part at NLL
\begin{equation}
T^{\rm NLL}(\gamma) = -\frac{1}{2}\asb\chi_0(\gamma)\frac{\partial \chi_0}{\partial \gamma}\sim \frac{\asb}{2\gamma^3}-\frac{\asb}{2(1-\gamma)^3}\; .
\label{NLL Scale changing}
\end{equation}
This scale change (\ref{NLL Scale changing}) is in agreement with the difference between the leading poles of symmetric and asymmetric NLL kernels \eqref{eq:chi0s} and \eqref{eq:chi0a}
\begin{align}
&\chi_1^S\sim -\frac{1}{2\gamma^3}-\frac{1}{2(1-\gamma)^3}\; ,\\
&\chi_1^A\sim\frac{0}{2\gamma^3}-\frac{1}{(1-\gamma)^3} \;.
\end{align}
In NNLL order, we have expand (\ref{SYM to AYSM}) to the second order in $\omega$ of $\chi_0$ and to the first order in $\omega$ in $\chi_1$, which gives
\begin{equation}
\chi_0(\gamma)-\frac{\omega}{2}\frac{\partial \chi_0}{\partial \gamma}+\frac{1}{2}\left(\frac{\omega}{2}\right)^2\frac{\partial^2 \chi_0}{\partial \gamma^2}+\asb\chi_1(\gamma)-\asb\frac{\omega}{2}\frac{\partial \chi_1}{\partial \gamma}.
\label{expansion to NNLL}
\end{equation}
Now in order to find scale changing terms at NNLL accuracy, we need to keep terms in the solution for the $\omega$ at least up to NLL i.e.
\begin{equation}
\omega=\asb\left[\chi_0\left(\gamma-\frac{\omega}{2}\right)+\asb\chi_1\left(\gamma-\frac{\omega}{2}\right)\right]\approx \asb\chi_0(\gamma)-\frac{1}{2}\asb^2\chi_0\frac{\partial \chi_0}{\partial \gamma}+\asb^2\chi_1.
\end{equation}
Plugging this expansion of $\omega$ into (\ref{expansion to NNLL}) and keeping terms up to NNLL, we get for the scale changing transformation at this order (see also \cite{Marzani:2007gk})
\begin{equation}
T^{\rm NNLL}(\gamma)=-\frac{1}{2}\frac{\partial \chi_0}{\partial \gamma}\left(-\frac{1}{2}\asb^2\chi_0\frac{\partial \chi_0}{\partial \gamma}\right)-\frac{1}{2}\asb^2\chi_1\frac{\partial \chi_0}{\partial \gamma}+\frac{1}{8}(\asb\chi_0)^2\frac{\partial^2 \chi_0}{\partial \gamma^2}-\frac{1}{2}\asb^2\chi_0\frac{\partial \chi_1}{\partial \gamma} \; .
\end{equation}
The  pole structure of this scale changing transformation at  $\gamma=0$ and $\gamma=1$ is 
\begin{align}
&T(\gamma)\sim-\frac{1}{2\gamma^5}+\frac{0}{\gamma^4}+{\cal O}(\frac{1}{\gamma^3})\; ,\\
&T(\gamma)\sim\frac{3}{2(1-\gamma)^5}+\frac{0}{(1-\gamma)^4}+{\cal O}(\frac{1}{(1-\gamma)^3})\; .
\label{scale changing in NNLL}
\end{align}
This result is the same whenever we are using $\chi_1$ either from exact N=4 sYM NLL kernel or the NLL kernel derived from $\omega$ expansion.  We see that the scale changing does not introduce any subleading poles $1/\gamma^4,1/(1-\gamma)^4$. It can be easily verified that the scale changing at  NNLL (\ref{scale changing in NNLL}) is consistent with the difference of the leading poles of NNLL kernels obtained from expanding \eqref{eq:chi0a} and \eqref{eq:chi0s}
\begin{align}
&\chi_2^A\sim \frac{0}{2\gamma^5}+\frac{2}{(1-\gamma)^5} \; ,\\
&\chi_2^S\sim \frac{1}{2\gamma^5}+\frac{1}{2(1-\gamma)^5} \; .
\end{align}

\subsection{Leading and subleading poles in  expansion of kernel with kinematical constraint}

There is an interesting feature of the subleading terms   extracted from the  expansion 
in $\omega$ of the $\chi$ function with $\omega$ shift. It is expected that a $\chi_k$  has the leading pole $\sim \frac{1}{\gamma^{2k+1}}$. We also observed that the subleading pole $\sim \frac{1}{\gamma^{2k}}$  vanish at NLL and NNLL order both in $N=4$ sYM results and for the kernel with kinematical constraint. This pattern can be shown to exist for kernel with kinematical constraint at higher orders and can be proved via mathematical induction. Let us assume that the kernel $\chi_{k}$ possesses the above discussed properties, and we will prove that the higher order kernel $\chi_{k+1}$ also has the same properties.
Let us write the solution for the  BFKL equation in Mellin space up to $k+1$'th power in $\asb$ (this is indicated by the subscript on $\omega$)
\begin{equation}
\omega_{k}=\asb\left[\chi_0+\asb\chi_1+\asb^2\chi_2+...+\asb^k \chi_k\right] \; .
\end{equation}
Now let us assume that the kinematical constraint introduces some $\omega$ dependence into the kernel eigenvalue so that the BFKL eigenvalue equation can be written as
\begin{equation}
\omega = \asb \chi^{\omega}(\gamma) \, ,
\end{equation}
and let us 
expand the eigenvalue  in $\omega$  and keep  terms to find $\omega_{k+1}$

\begin{equation}
\omega_{k+1}=\asb\left[\chi^{(0)}+\chi^{(1)}\omega_k+\frac{1}{2!}\chi^{(2)}\omega_k^2+...+\frac{1}{k!}\chi^{(k)}\omega_k^k+\frac{1}{(k+1)!}\chi^{(k+1)}\omega_k^{k+1}\right] \; ,
\label{whole}
\end{equation}
with the \(\chi^{(i)}\) is the \(i\)'th derivative of \(\chi^{\omega}\) on \(\omega\) set at $\omega=0$ and of course $\chi_0=\chi^{(0)}$. Let us now extract the \(\asb^{k+2}\) term from (\ref{whole}). We note, that one needs to keep terms up to $\omega_k$ on the r.h.s which will contribute to $\omega_{k+1}$.

For an arbitrary term in the $[\dots]$ bracket of (\ref{whole}) we have
\begin{equation}
\frac{\chi^{(i)}}{i!}\omega_k^i=\frac{\asb^i \chi^{(i)}}{i!}\left(\chi_0+\asb\chi_1+\asb^2\chi_2+\dots+\asb^k \chi_k\right)^i\;.
\label{arbitratyterm}
\end{equation}
In order to find the \(\asb^{k+2}\) term in \eqref{whole}, we need the \(\asb^{k+1-i}\) term in 
\begin{equation}
\left(\chi_0+\asb\chi_1+\asb^2\chi_2+\dots+\asb^k \chi_k\right)^i \; .
\label{reduced}
\end{equation}
The general expansion of (\ref{reduced}) could be really complicated. But ignoring the coefficients, we can express an arbitrary term in (\ref{reduced}) in the following form
\begin{equation}
\prod_{l=0}^{k}\chi_l^{j_l}=\chi_0^{j_0}\chi_1^{j_1}...\chi_k^{j_k} \, ,
\label{combination}
\end{equation}
with a constraint on powers  \(\{j_l\}\)
\begin{equation}
\sum_{l=0}^{k}j_l=i.
\label{constaint1}
\end{equation}
Since a \(\chi_l\) term would automatically bring a power of \(\asb^l\), if we demand (\ref{combination}) to have  the term of order  \(\asb^{k+1-i}\), we have to put a new constraint on \(\{j_l\}\), which is
\begin{equation}
\sum_{l=0}^{k}j_l l=k+1-i\, .
\label{constaint2}
\end{equation}
Thus, for a  term (\ref{combination}), given that the leading pole of $\chi_l\sim \frac{1}{\gamma^{2l+1}}$ and under the constraints (\ref{constaint1}), (\ref{constaint2}), its leading pole can be given as
\begin{equation}
\prod_{l=0}^{k}\left(\frac{1}{\gamma^{2l+1}}\right)^{j_l}=\left(\frac{1}{\gamma}\right)^{\sum_{l=0}^{k}(2l+1)j_l}=\left(\frac{1}{\gamma}\right)^{(2k+2-i)} \; .
\end{equation}
Also, knowing that every \(\chi_l\) in (\ref{combination}) doesn't have a subleading pole, we can see that the subleading pole of this \(\asb^{k+1-i}\) term will also vanish. 

Now, we can get back to expression (\ref{arbitratyterm}). Knowing  that the derivative \(\chi^{(i)}\) has the leading pole \(\sim{\gamma^{-(1+i)}}\), and a vanishing subleading pole, we can conclude that the leading pole in \(\chi_{k+1}\)  is \(\sim\gamma^{-(2k+3)}\), and its \(\gamma^{-(2k+2)}\) pole vanishes. 

	\section{Properties of the kernel in Mellin space}
	\label{sec:PropsEigenvs}
In this section we shall investigate in detail the differences between the different constraints on the level of the Mellin transform.

\begin{figure}[!htp]
	\centering
	\subfigure{
		\includegraphics[width=0.48\linewidth]{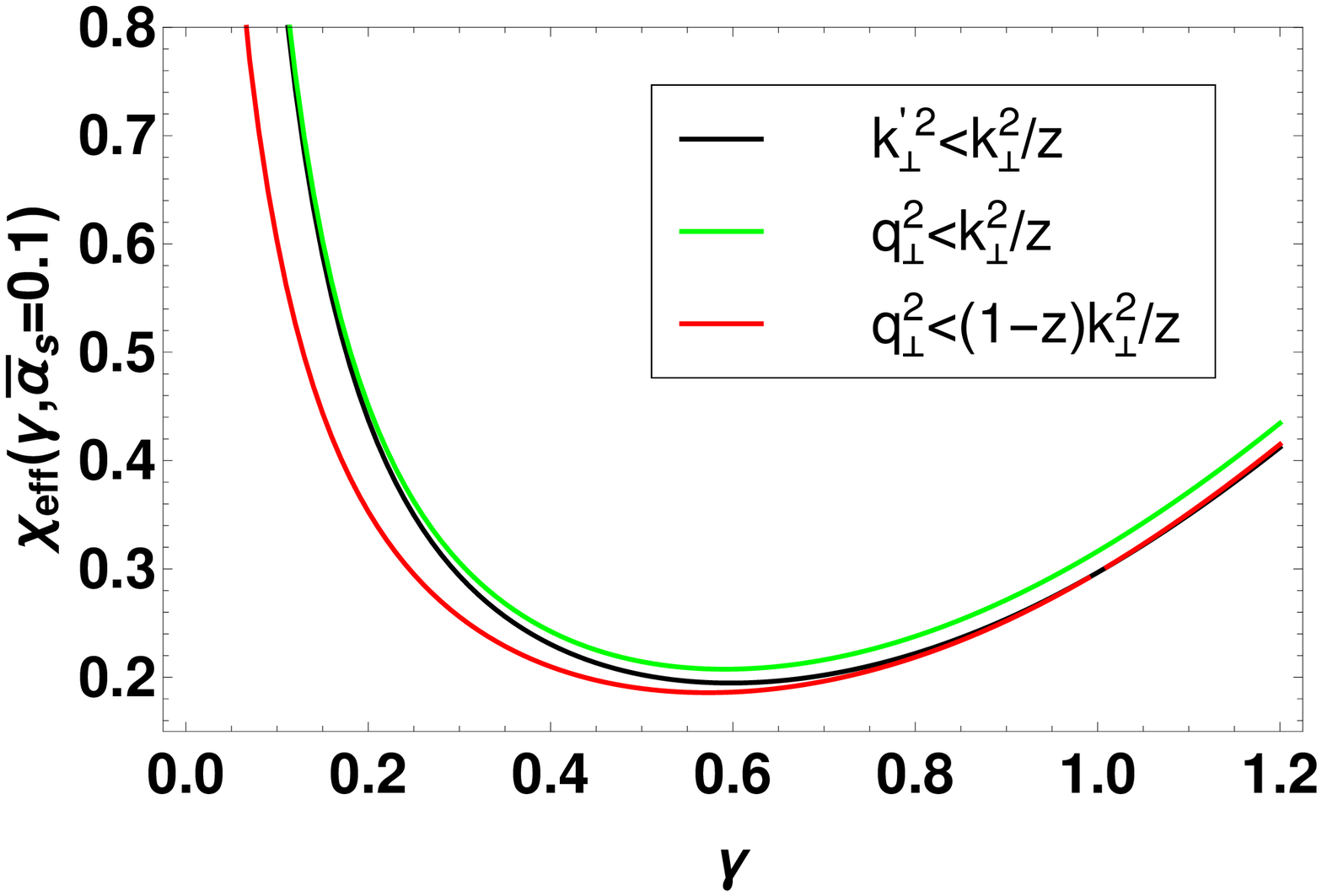}
	}
	\subfigure{
		\includegraphics[width=0.48\linewidth]{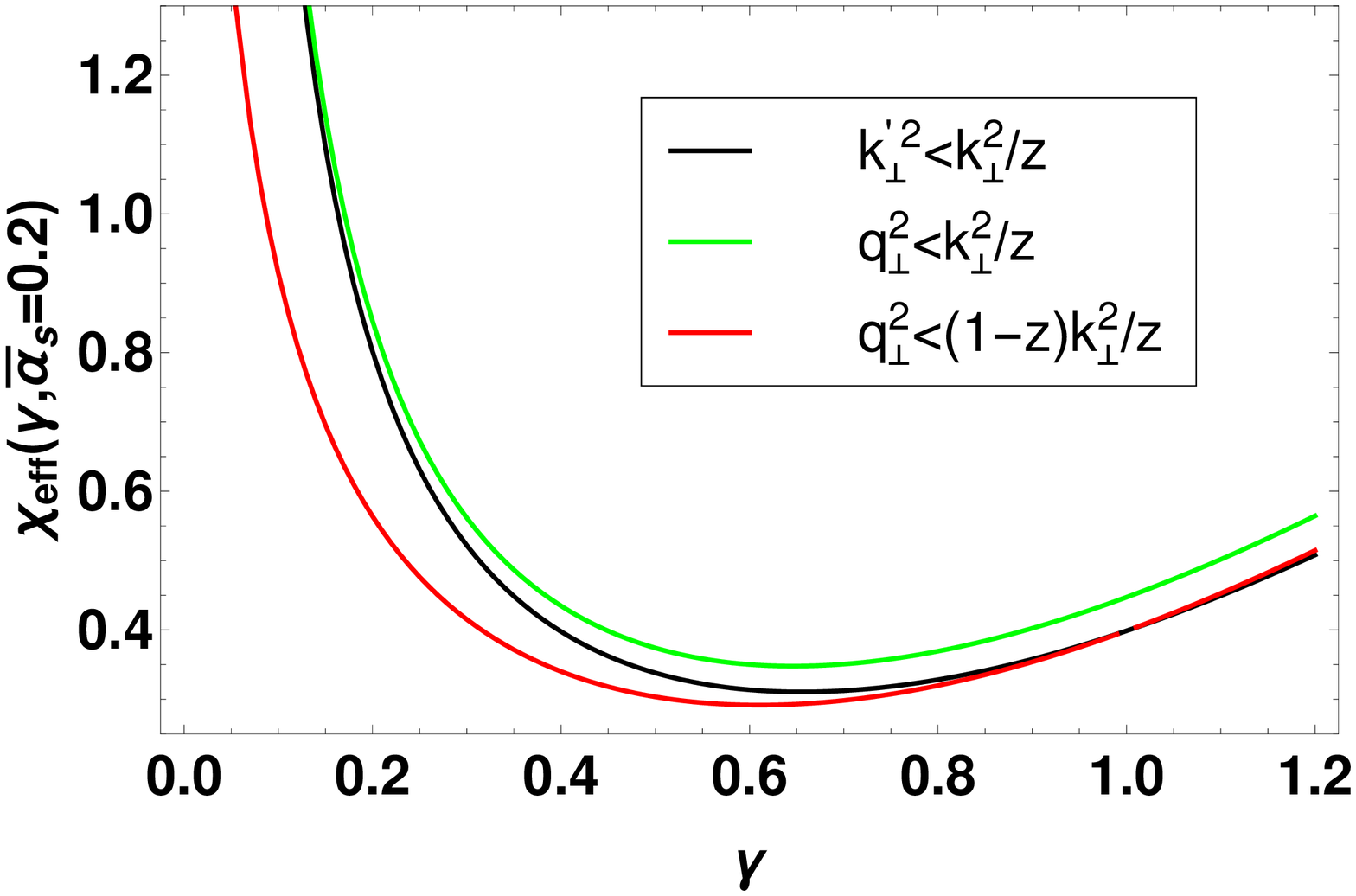}
	}
	\caption{The effective Mellin transformed kernel $\chi_{\rm eff}$ for the three different versions of the  kinematical constraint  \eqref{eq:kincon-f},\eqref{eq:kincon-jk},\eqref{eq:kinconorig} as a function of Mellin variable $\gamma$. Left: $\asb=0.1$, right plot $\asb=0.2$.}
	\label{fig:chigammadep1}
\end{figure}

The BFKL equation in momentum space with  the  constraint \eqref{eq:kincon-f}  is be given by
\begin{multline}
\label{eq:bfkl_fkc}
\cF(x,\kt^2)=\cF^{(0)}(x,\kt^2)\\
+\asb\int_{x}^{1}\frac{dz}{z}\int\frac{d^2 \bqt}{\pi \qt^2}\left[\cF\left(\frac{x}{z},
\left|\bkt+\bqt\right|^2\right)\Theta\left(\kt^2-\frac{z}{1-z}\qt^2\right)-\Theta(\kt^2-\qt^2)\cF\left(\frac{x}{z},\kt^2\right)\right] \; ,
\end{multline}
where the kinematical constraint is implemented onto the real emission term in the form of the Heaviside function. 
Performing the  Mellin transformation of Eq.~\eqref{eq:bfkl_fkc} into the  \(\omega\) space 
 gives
\begin{multline}
\bar{{\cal F}}(\omega,\kt^2)=\bar{\cal F}^{(0)}(\omega,\kt^2)\\
+\asb\int_{x}^{1}\frac{dz}{z}z^{\omega}\int\frac{d^2 \bqt}{\pi \qt^2}\left[\bar{{\cal F}}\left(\omega,
\left|\bkt+\bqt\right|^2\right)\Theta\left(\kt^2-\frac{z}{1-z}\qt^2\right)-\Theta(\kt^2-\qt^2)\bar{{\cal F}}\left(\omega,\kt^2\right)\right] \; .
\end{multline}
Performing another Mellin transformation, this time on $\kt^2$ variable, 
we get the following algebraic form of the BFKL equation
\begin{equation}
\tilde{{\cal F}}(\omega,\gamma)=\tilde{{\cal F}}^{(0)}(\omega,\gamma) + \frac{\asb}{\omega} \chi(\gamma,\omega) \tilde{{\cal F}}(\omega,\gamma) \, ,
\end{equation}
where the kernel is 
\begin{equation}
 \chi(\gamma,\omega) \; = \; \int \frac{d^2 \bqt}{\pi   \qt^2}\left[\left(1+\frac{\qt^2}{\kt^2}\right)^{-\omega}\left(\frac{\left|\bkt + \bqt\right|^2}{\kt^2}\right)^{\gamma-1}-\Theta(\kt^2-\qt^2)\right]\, .
 \label{eq:chi_fkc1}
\end{equation}
One can further simplify this expression by performing the change of variables
 \(u=q_T^2/k_T^2\). This gives also
\begin{equation}
\frac{\left|\bkt+\bqt\right|^2}{\kt^2}=1+u+2\sqrt{u}\cos \phi\; ,
\end{equation}
where $\phi$ is defined as the angle between $\bqt$ and $\bkt$ vectors.  Then the kernel \eqref{eq:chi_fkc1} becomes
\begin{equation}
\chi(\omega,\gamma)=\int_{0}^{2\pi}\frac{d\phi}{2\pi}\int_{0}^{1}\frac{du}{u}\left[(1+u)^{-\omega}\left(1+u^{\omega+1-\gamma}\right)(1+u+2\sqrt{u}\cos \phi)^{\gamma-1}-1\right] \, .
\end{equation}
We note that the integration is from $0$ to $1$ here, and the term $u^{\omega+1-\gamma}$ arises when one integrates over the region $1<u<\infty$ after the transformation of $u$ to $1/u$.
The angular integral can be performed, resulting in the  hypergeometric function,
\begin{equation}
 \label{eq:chi_fkc2}
\chi(\omega,\gamma)=\int_{0}^1\frac{du}{u}\left[(1+u)^{-\omega}\left(1+u^{\omega+1-\gamma}\right){_2F_1}(1-\gamma,1-\gamma;1;u)-1\right]\;.
\end{equation}
For comparison, the BFKL kernel with the constraint \(k_T^2>zq_T^2\), or Eq.\eqref{eq:kincon-jk} the low $z$ approximation, is given by \cite{Kwiecinski:1996td}
\begin{equation}
 \label{eq:chi_fkc2a}
\chi(\omega,\gamma)=\int_{0}^1\frac{du}{u}\left[\left(1+u^{\omega+1-\gamma}\right){_2F_1}(1-\gamma,1-\gamma;1;u)-1\right] \; .
\end{equation}
The leading pole position is of course given by the solution to 
\begin{equation}
\omega = \asb \chi(\omega,\gamma) \; .
\label{eq:chi_eq}
\end{equation}
This transcendental equation cannot be solved analytically with the complicated $\omega$ dependence inside the kernel, nevertheless can be solved formally to give an `effective' kernel
\begin{equation}
\omega= \chi_{\rm eff}(\gamma,\asb) \; .
\end{equation}
In order to obtain the $\chi_{\rm eff}$ we have solved the Eq.\eqref{eq:chi_eq} numerically for all three versions of the kernel  \eqref{eq:shifted_asym},\eqref{eq:chi_fkc2},\eqref{eq:chi_fkc2a} corresponding to kinematical constraints,
 \eqref{eq:kincon-f},\eqref{eq:kincon-jk},\eqref{eq:kinconorig}. 
The results of the calculation for two different values of the strong coupling constant, $\asb=0.1,0.2$ are shown in  Fig. \ref{fig:chigammadep1}.

The first feature of the effective kernels in all cases is that they are strongly modified in the large $\gamma$ region. The results from the constraint \eqref{eq:kincon-f} and \eqref{eq:kinconorig} are very close in the $\gamma \simeq 1$ region. Comparing Eq.~\eqref{eq:chi_fkc2}
and \eqref{eq:shifted_asym} we found that they are in fact identical at $\gamma=1$  for any value of  $\omega$. One can observe that for $z\rightarrow 1$ both \eqref{eq:kinconorig} and \eqref{eq:kincon-f} lead to a strong suppression of the anti-collinear phase-space, i.e. $\kt^2 < \kt^{'2}$,  whereas even for $z\simeq 1$ the constraint \eqref{eq:kincon-fa} leaves some phase space for integration.  On the other  hand the constraints \eqref{eq:kincon-jk} and \eqref{eq:kinconorig} are very close in the small $\gamma$ region while the \eqref{eq:kincon-f} is lower in that range. This can be understood from the fact that in the collinear region,
i.e. for $\kt^2 > \kt^{\prime 2}$ the latter constraint is stronger, particularly if $z$ is large.

Next, we have checked numerically the coefficients in front of the poles in $\gamma$ and found agreement between different forms of constraints  in the leading $\sim 1/(1-\gamma)^{2k+1}$ and subleading poles $\sim 1/(1-\gamma)^{2k}$ poles when the  expressions for the kernels are expanded out to NLL and NNLL .

There is one notable difference between \eqref{eq:kincon-f} and the other two constraints, in that in the first one the collinear pole in $\gamma$ is absent. This can be understood as for large values of $z>1/2$ constraint \eqref{eq:kincon-f}
implies the restriction of the integration over collinear region as well.

\begin{figure}[!htp]
	\centering
	\includegraphics[width=0.45\linewidth, height=0.225\textheight]{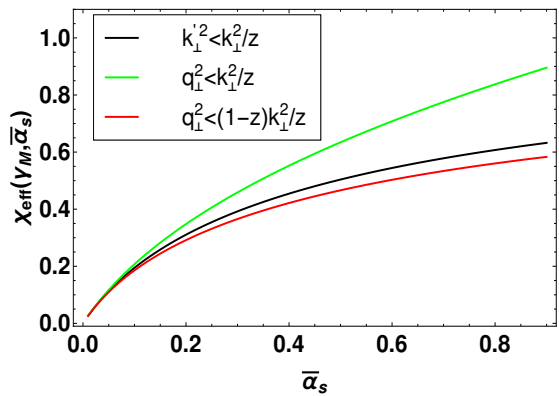}
	\caption{The values of the intercept  $\omega_{I\!\!P}= \chi_{\rm eff}(\gamma_M,\asb)$, Eq.~\eqref{eq:intercept}, as a function of ${\bar\alpha}_s$ for different choices of the kinematical constraint.}
	\label{fig:chiasgam}
\end{figure}

From $\chi_{\rm eff}$ one can find the leading behavior as $x\rightarrow 0$. The intercept which is found from
\begin{equation}
\frac{d\chi_{\rm eff}(\gamma,\asb)}{d\gamma}|_{\gamma=\gamma_M}=0 \, , \quad \omega_{I\!\!P}= \chi_{\rm eff}(\gamma_M,\asb) \; ,
\label{eq:intercept}
\end{equation}
and is plotted in Fig.~\ref{fig:chiasgam}.  All the constraints give similar intercept for values of $\asb$ up to about $0.2$. The constraint $\eqref{eq:kincon-jk}$ gives somewhat larger values, when $\asb$ is further increased, while the two constraints $\eqref{eq:kinconorig}$ and $\eqref{eq:kincon-f}$ give comparable values for the intercepts even
for very large values of the strong coupling. This is consistent with the shape of the kernel discussed  before.


	\section{Differential form of the evolution equation with kinematical constraints}
	\label{sec:DiffForm}
	
	The evolution equation in the presence of the kinematical constraint becomes an integral equation, over the longitudinal variable $z$ as is evident from Eq.~\eqref{eq:bfkl_fkc}. Equations in this form were solved in the momentum space via different techniques in the past. One technique utilized in the past in Ref.~\cite{Kwiecinski:1997ee} was based on the extension of the method introduced in \cite{Kwiecinski:1990gc}. This method incorporates the interpolation in two variables $x$ and $\kt^2$ with orthogonal polynomials. By doing this the  integral equations can be transformed into the set of linear algebraic equations and solved by simply inverting the large matrix. An alternative method was used in \cite{Ciafaloni:2003rd} and it was checked that it coincides with the one proposed in \cite{Kwiecinski:1997ee}.

In this paper we shall propose and utilize another method which reduces the integral equation to the differential one.
	The differentiation of the BFKL equation by $x$ is motivated by the reduction of computational complexity of the integral on the right hand side of the equation. After performing the derivative we find, that no matter what is the form of the kinematical constraint the resulting equation can be written in the differential form and  has the same form containing two terms - real emission term with shifted argument $x$ and virtual correction term - under the remaining integral over the emitted momentum ${\bf q}$. The argument of the real emission term is shifted such, that if  the kinematical constraint takes the form $\theta\left(k_C\left({\bf q},{\bf k}\right)-z\right)$, then the argument of the unintegrated gluon density function in the real emission term changes in following way:
\begin{equation}
x\rightarrow x\;\max\left\{1,\frac{1}{k_C\left({\bf q},{\bf k}\right)}\right\}\;.
\end{equation}
Such reformulation of the BFKL equation with kinematical constraint was first proposed in \cite{Andersson:1995ju}, and similar ideas were also discussed in \cite{Beuf:2014uia} in the context of the Balitsky-Kovchegov evolution equation in transverse coordinate space.
Below we derive the differential form of the BFKL equation for each case of the kinematical constraint. 


Let us begin by differentiating in $x$ the equation with kinematical constraint in the low $z$ approximation, Eq.~\eqref{eq:kincon-jk}.
The equation formally differentiated in $x$  reads 
\begin{equation}\label{eq:unresmBFKLskcdiff}
\begin{split}
\frac{\partial{ {\cal F}}(x,{\kt^2})}{\partial x}=\frac{\partial{ {\cal F}}_0(x,{\kt^2})}{\partial x}&+{\bar\alpha_s}\frac{\partial}{\partial x}\int\limits_{x}^{1}\frac{dx^\prime}{x^\prime}\;\int\frac{d^2{\bqt}}{\pi{{\qt}^2}}\bigg\{{ {\cal F}}\left(x^\prime,|{\bkt}+{\bqt}|^2\right)\Theta\left(\kt^2-\frac{x}{x^\prime}\qt^2\right) \\ & -\,\Theta\left(\kt^2-\qt^2\right){ {\cal F}}\left(x^\prime,{\kt^2}\right)\bigg\}\;.
\end{split}
\end{equation}
where we performed the change of variable $z\rightarrow x^\prime=x/z$.
After performing the derivative we get:
\begin{equation}\label{eq:unresmBFKLskcdiffR}
\begin{split}
\frac{\partial{ {\cal F}}(x,{\kt^2})}{\partial x}=&\frac{\partial{ {\cal F}}_0(x,{\kt^2})}{\partial x}-\frac{{\bar\alpha_s}}{x}\;\int\frac{d^2{\bqt}}{\pi{{\qt}^2}}\bigg\{{ {\cal F}}\left(x\,\max\left(1,\frac{\qt^2}{\kt^2}\right),|{\bkt}+{\bqt}|^2\right)\Theta\left(\frac{\kt^2}{\qt^2}-x\right) \\-&\,\Theta\left(\kt^2-\qt^2\right){ {\cal F}}\left(x,{\kt^2}\right)\bigg\}\;,
\end{split}
\end{equation}

where the expression $x\,\max\left(1,\frac{q_T^2}{k_T^2}\right)$ emerges from two terms generated by the derivative acting on the integral on the right hand side of the~\eqref{eq:unresmBFKLskcdiffR}. The first one coming from derivative on the boundary of the integral generating term containing ${\cal F}\left(x,|{\bkt}+{\bqt}|^2\right)\Theta\left({\kt^2}-{\qt^2}\right)$ and the second one acting on the $\Theta$-function from kinematical constraint generating a term containing ${ F}\left(x\frac{\qt^2}{\kt^2},|{\bkt}-{\bqt}|^2\right)\Theta\left({\qt^2}-{\kt^2}\right)$ - which combine giving the result in~\eqref{eq:unresmBFKLskcdiffR}.

Rewriting the derivative as a derivative by $\ln{1/x}$ instead of $x$ results in
\begin{equation}\label{eq:unresmBFKLskcdiffR}
\begin{split}
\frac{\partial{ {\cal F}}(x,{\kt^2})}{\partial\ln{1/x}}=&\frac{\partial{ {\cal F}}_0(x,{\kt^2})}{\partial\ln{1/x}}+{\bar\alpha_s}\;\int\frac{d^2{\bqt}}{\pi{{\qt}^2}}\bigg\{{ {\cal F}}\left(x\,\max\left(1,\frac{\qt^2}{\kt^2}\right),|{\bkt}+{\bqt}|^2\right)\Theta\left(\frac{\kt^2}{\qt^2}-x\right)
\\-&\,\Theta\left(\kt^2-\qt^2\right){ {\cal F}}\left(x,{\kt^2}\right)\bigg\}\;.
\end{split}
\end{equation}

The case of constraint $\kt'^2 < \kt^2/z$ can be treated in exactly the same way with the result
 \begin{equation}\label{eq:unresmBFKLpkcdiffR}
 \begin{split}
 \frac{\partial{ {\cal F}}(x,{\kt^2})}{\partial\ln{1/x}}=&\frac{\partial{ {\cal F}}_0(x,{\kt^2})}{\partial\ln{1/x}}+{\bar\alpha_s}\;\int\frac{d^2{\bqt}}{\pi{{\qt}^2}}\bigg\{{ {\cal F}}\left(x\,\max\left(1,\frac{\kt^{'2}}{\kt^2}\right),|{\bkt}+{\bqt}|^2\right)\Theta\left(\frac{\kt^2}{\kt^{'2}}-x\right)
 \\-&\,\Theta\left(\kt^2-\qt^2\right){ {\cal F}}\left(x,{\kt^2}\right)\bigg\}\;.
 \end{split}
 \end{equation}
 Thus we see very intersting result in that the entire effect of the kinematical constraint is to shift the longitudinal variable in the argument of the unintegrated gluon density in the real term by a factor which includes a combination of ratio of  transverse momenta.

The case with the kinematical constraint $\qt^2 < \frac{1-z}{z} \kt^2$ is a bit different, since
in this case the term originating from the derivative acting on the integral boundary is equal to $0$. This is because of the $\Theta$-function being evaluated at $-\infty$. In other words, this constraint means that $x'>x \frac{k_T^2+q_T^2}{k_T^2}$ which is always stronger than $x'>x$ and therefore the boundary term will not contribute here. This results in the following differential equation

\begin{equation}\label{eq:unresmBFKLkcdiffR}
\begin{split}
\frac{\partial{ {\cal F}}(x,{\kt^2})}{\partial\ln{1/x}}=\frac{\partial{{\cal F}}_0(x,{\kt^2})}{\partial\ln{1/x}}&+{\bar\alpha_s}\int\frac{d^2{\bqt}}{\pi{{\qt}^2}}\bigg\{{{\cal F}}\left(x\,\frac{\kt^2+\qt^2}{\kt^2},|{\bkt}+{\bqt}|^2\right)\Theta\left(\frac{\kt^2}{\qt^2+\kt^2}-x\right)\\ & -\,\Theta\left(\kt^2-\qt^2\right){{\cal F}}\left(x,{\kt^2}\right)\bigg\}\;.
\end{split}
\end{equation}

	\section{Numerical method and results}
	\label{sec:Numerical}


\subsection{Numerical results}
	In this section we shall present the numerical results for the solution of the BFKL equation with different forms of the kinematical constraint. The details for the  numerical method are described in the Appendix.  The equations are solved using the boundary condition

\begin{equation}
{\mathcal F}\left(x,k_T\right)=\exp\left[p_0\log{(k_T/\mu)}^2+p_1\log^2{(k_T/\mu)}^2\right]\left(1-x\right)^{p_2}\;,
\end{equation}
with $p_0=p_1=-0.1$ and $p_2=2$ and $\mu=1 \, {\rm GeV}$.
Since we are not doing any phenomenology here this boundary condition has not being fitted to any data and is for illustrative purposes only.
The particular choice of $k_T$ dependence is motivated by analitical solution of the LO BFKL equation and guarantes fast convergence of the solution. The term $(1-x)^{p_2}$ guaantees that the gluon is suppressed for large values of $x$ so that the integral over the kernel can be extended to large $x$ values. 
We introduce the lower cutoff onto the transverse momenta $Q_0=0.1 \; {\rm GeV}$.  The results were obtained for the fixed value of the coupling constant $\asb=0.2$.

\begin{figure}[h]
	\centering
	\subfigure{
		\includegraphics[width=0.45\linewidth, height=0.225\textheight]{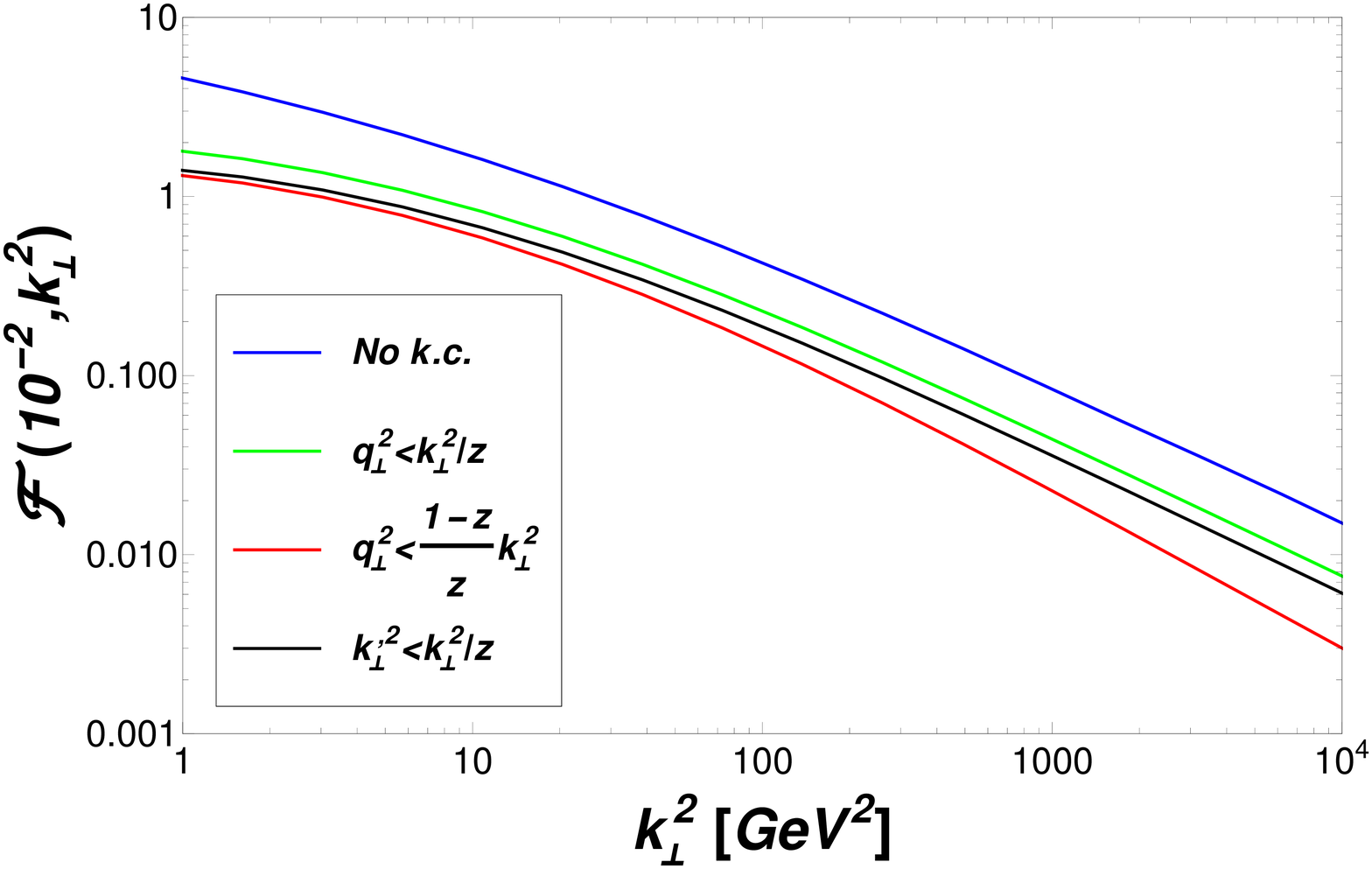}
	}
	\subfigure{
		\includegraphics[width=0.45\linewidth, height=0.225\textheight]{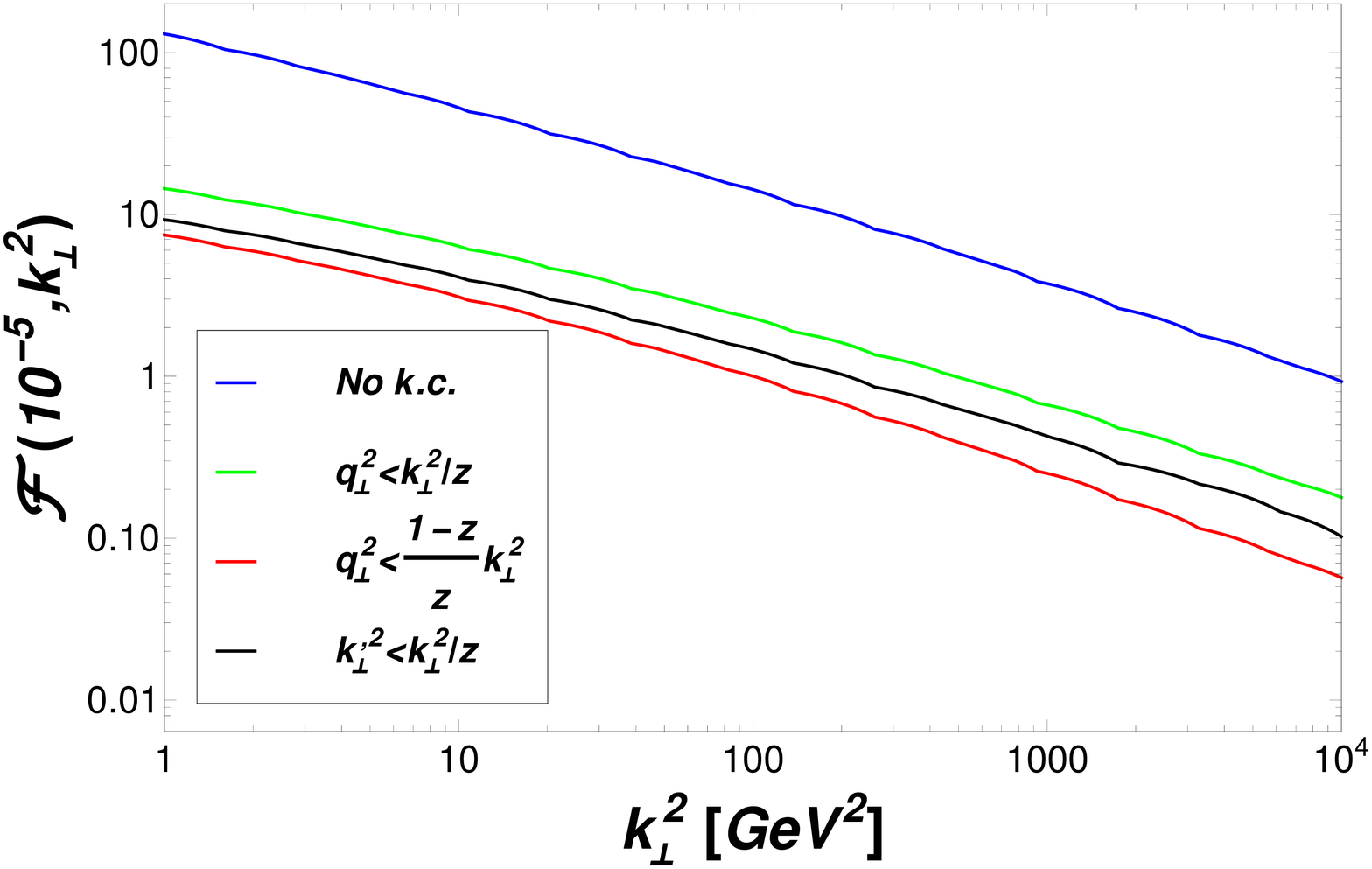}
	}
	\caption{Solutions of the BFKL equations with different forms of the kinematical constraint as functions of $k_T$ for constant ${\bar\alpha}_S=0.2$. Left plot: $x=10^{-2}$, right plot: $x=10^{-5}$.}
	\label{fig:k2-1}
\end{figure}

\begin{figure}[h]
	\centering
	\subfigure{
		\includegraphics[width=0.45\linewidth, height=0.225\textheight]{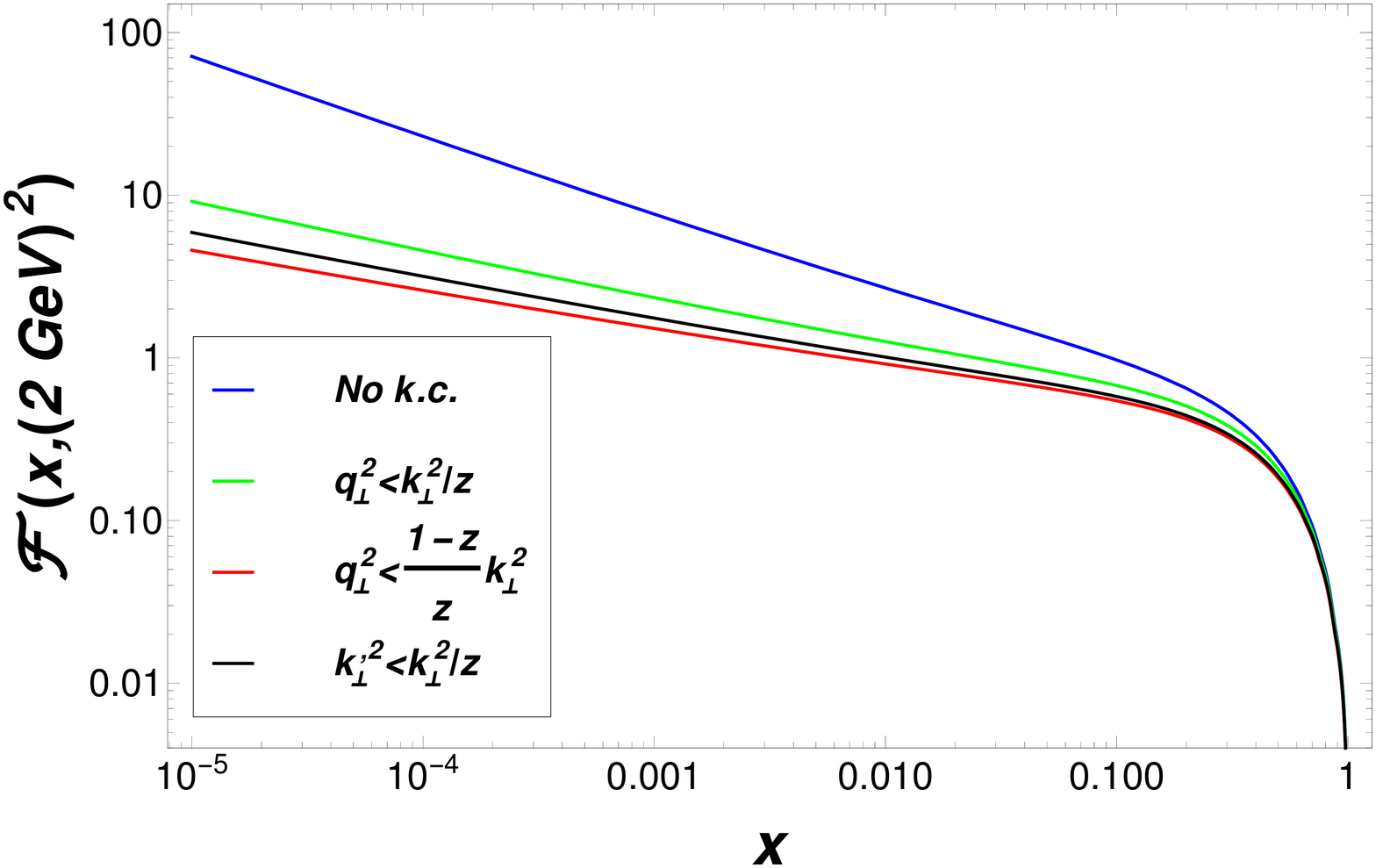}
	}
	\subfigure{
		\includegraphics[width=0.45\linewidth, height=0.225\textheight]{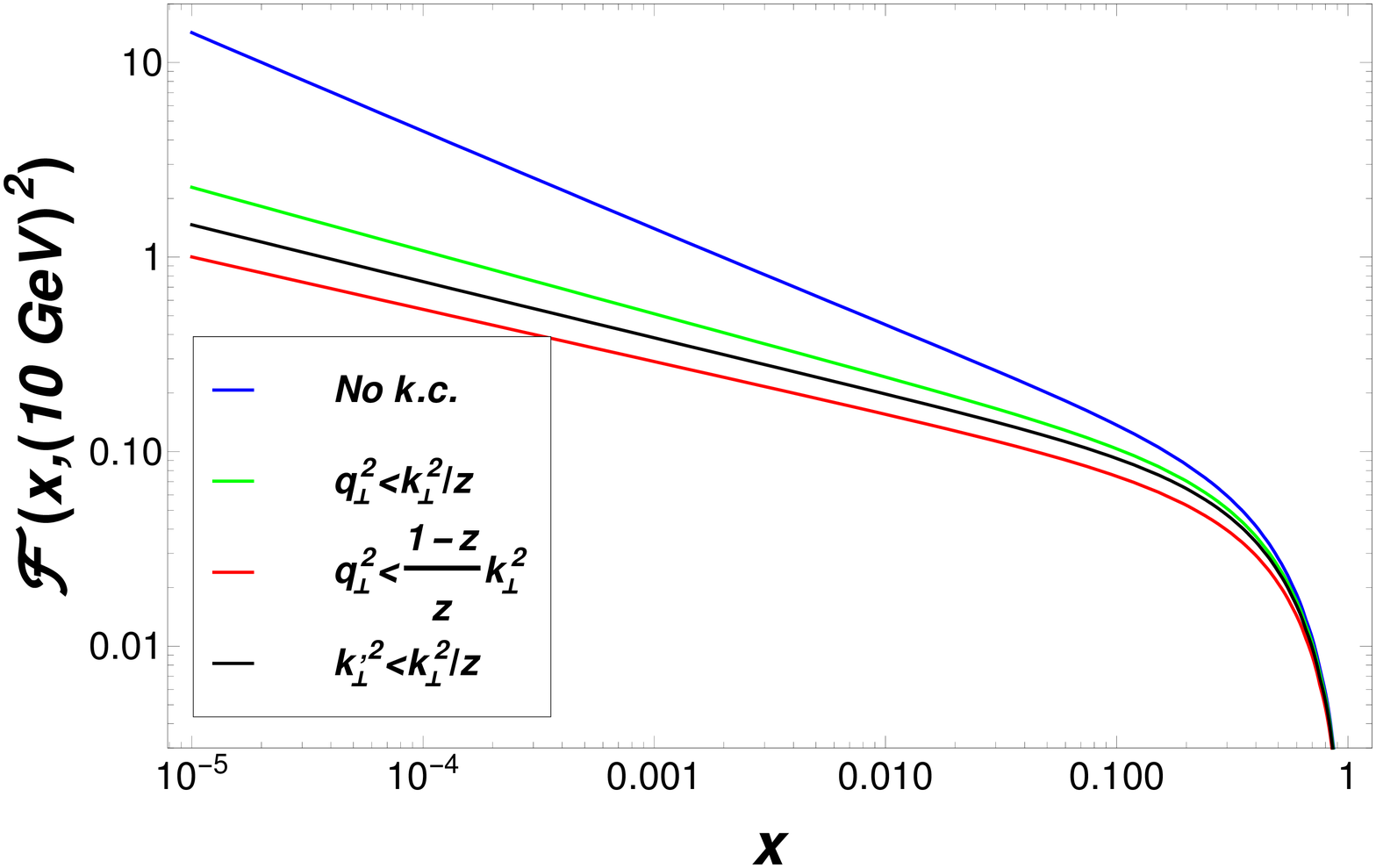}
	}
	\caption{Solutions to the BFKL equation with three different forms of the constraint as a function  of $x$ for constant ${\bar\alpha}_S=0.2$. Left plot: $\kt=2\,  {\rm GeV}$, right plot $\kt=10 \, {\rm GeV}$.}
	\label{fig:x-1}
\end{figure}

The results for the solution as a function of the $\kt^2$ for fixed values of $x$ and different forms of the kinematical constraint are shown  in plots in Figs.~\ref{fig:k2-1} together with the solution to the LL BFKL.
 The results reflect the trend of the semi-analytical results from the section~\ref{sec:PropsEigenvs}.  In plots of ${\kt}^2$ dependence for $x=10^{-2}$ it is clear that the solution with the full kinematical constraint and the solution of the equation with the kinematical constraint in the $\kt^{\prime 2}<k_T^2/z$ approximation get close to each other for low ${k_T}^2$.
 On the other hand, for large values of $\kt^2$ it is the two constraints (\ref{eq:kincon-jk}) and (\ref{eq:kincon-f}) which 
 come close to each other and (\ref{eq:kincon-f}) has a different behavior. This is related to the different behavior in the collinear limit, i.e. the fact that constraint (\ref{eq:kincon-f}) introduces modification in the collinear limit as well as discussed previously. In any case the solution with the constraint (\ref{eq:kincon-jk}) always leads to the results which is largest, consistent with the analysis from the Mellin space. Similar conclusions can be reached from the analysis  of the plots   with the $x$ dependence for small ${k_T}^2$ shown in Figs.\eqref{fig:x-1}. The small $z$ approximation of the kinematical constraint results in the solution being  closer to the solution of the equation with the ${q_T}^2\simeq{k^{\prime}_T}^2$ approximation of the kinematical constraint for large ${k_T}^2$.
From Fig.\eqref{fig:x-1}  we see the onset of the power behaviour in $x$, and in all versions of the kinematical constraint the intercept is strongly reduced with respect to the LL approximation. The differences between the solutions with different versions of the kinematical constraint are much smaller than between resummed and LL result. Nevertheless, the differences between different versions of the kinematical constraint are non-negligible and can reach up to a factor 2-3 depending on the range of $x$ and $\kt$.

	\section*{Conclusions}
In the paper we have performed a detailed analysis of the  three versions of kinematical constraints imposed onto the momentum space BFKL equation. We used the fixed coupling BFKL equation as a basic equation on which kernel's the constrains are imposed. This choice allowed us to investigate semi-analytically the properties of the kinematical constrains in the
Mellin space and to compare to known results in sYM version of the BFKL equations where the results are known up to NNLL order.
In particular, we observed that the leading poles  in Mellin space generated by the kinematical constraint obey maximal transcendentality principle. In addition we proved that the subleading poles vanish at NLL and NNLL order both for BFKL and its sYM version, and that such feature can be expected to hold to all orders. 
Furthermore we observed that the $\chi_{eff}$ function obtained with the most general version of the constraint i.e. (\ref{eq:kincon-f}) for large values of $\gamma$  agrees well with the  more commonly used version of the constraint i.e (\ref{eq:kinconorig}). However, at small values of $\gamma$ the $\chi_{eff}$ function with (\ref{eq:kinconorig}) agrees with (\ref{eq:kincon-jk}) and both of them differ from the case when constraint (\ref{eq:kincon-f}) is imposed. In particular, it is important to note that the last one changes the collinear structure of the kernel i.e. it limits $z$ to $z<1/2$. While on formal grounds this effect might be neglected in the small $z$ limit it has nonnegligible impact on the resulting gluon density. We demonstrate that by numerical solutions of the momentum space BFKL equation for unintegrated gluon density for all versions of the constraint as well as for the unmodified BFKL equation.
The understanding of action of the kinematical constraint across the whole region of $z$ will be particularly important for ongoing efforts to unify the BFKL and DGLAP evolution schemes \cite{Gituliar:2015agu,Hentschinski:2016wya,Hentschinski:2017ayz} where effects of kinematical constrains may play a crucial role.
Furthermore, the inclusion of the kinematical constraint is known to be relevant for BFKL phenomenology \cite{Kwiecinski:1997ee,Hentschinski:2013id,vanHameren:2014ala,Iancu:2015joa,vanHameren:2015uia,vanHameren:2019ysa,Phukan:2019kfy}. However, so far only the simplest version of it was used in the context of BFKL. It will be interesting to see how different versions of the constraint perform while a fit to $F_2$ data is done or what are the differences when the description of processes probing gluon density at very low $x$ like high energy neutrinos or very forward dijet data is attempted.
	\section*{Acknowledgments}
We thank Stanis\l{}aw Jadach for comments and discussions. We thank Radek Zlebcik for discussions and collaboration at the initial stages of the project. Krzysztof Kutak acknowledges the support of the COST {\it Action CA16201 Unraveling new physics at the LHC through the precision frontier}. This work  was supported by the  Department of Energy Grants No. DE-SC-0002145 and DE-FG02-93ER40771, as well as the National Science centre, Poland, Grant No.\ 2015/17/B/ST2/01838. 
		\section*{Appendix}
		\subsection*{Algebraic form of the equation suitable for numerical solution}
	
		In order to solve the equation we shall use the method of expansion in the set of Chebyshev polynomials
		in the $\kt$ variable. Using this interpolation we can expand a function on the interval $[-1,1]$ in the following way
		\begin{equation}\label{eq:numderivative}
		\begin{split}
		f\left(\zeta\right)=\sum\limits_{j=0}^{N-1}f\left(\zeta_j\right)c_j\left(\zeta,N\right)=\sum\limits_{j=0}^{N-1}f_j\,c_j\left(\zeta,N\right)\;,
		\end{split}
		\end{equation}
		
		where
		\begin{equation}
		c_j\left(\zeta,N\right)=\frac{2}{N}\sum\limits_{i=0}^{N-1}T_i\left(\zeta_j\right)T_i\left(\zeta\right)o_i\; ,
		\end{equation}
		
		with $T_i\left(\zeta\right)$ a Chebyshev polynomial of the $i$-th order, $\zeta_j$ are nodes of the Chebyshev polynomial of the $N$-th order,i.e. $\zeta_j=\cos((j+\frac{1}{2})\frac{\pi}{N})$ and finally $o_i=1/2$ for $i=0$ and $o_i=1$ for $i>0$.
		In the following, we  use the unintegrated gluon distribution functions which are averaged over the angle, i.e. the dependence is only on $\kt^2$.
		Let us first rewrite the equation~\eqref{eq:unresmBFKLkcdiffR} using the expansion of the function ${ F}\left(x,{\kt^2}\right)$ in $\kt$ on a grid of Chebyshev nodes and in $x$ with trapezoidal interpolation
		\begin{equation}\label{eq:num1ststep}
		\begin{split}
		\frac{\partial{ {\cal F}}(x,{\kt}_i^2)}{\partial\ln{x}}\bigg|_{x=x_j}&=\frac{\partial{ {\cal F}}_0(x,{\kt}_i^2)}{\partial\ln{x}}\bigg|_{x=x_j}\\-\sum\limits_{m=0}^{N_k-1}\sum\limits_{n=0}^{N_x-1}{\bar\alpha_s}&\int\frac{d^2{\bqt}}{\pi{{\qt}^2}}\left\{{{\cal F}}\left(x_n,{\kt^2}_m\right)t\left(x_j\,\frac{{\kt^2}_i+\qt^2}{{\kt^2}_i},n\right)c_m\left(\zeta\left(|{\bkt}_i-{\bqt}|\right),N_k\right)\right.\\  &\qquad-\,\left.\Theta\left({\kt}^2_i-\qt^2\right){{\cal F}}\left(x_n,{\kt^2}_m\right)t\left(x_j,n\right)c_m\left(\zeta\left({\kt}_i\right),N_k\right)\right\}\;,
		\end{split}
		\end{equation}
		
		where ${\kt}_m$ are values projected from Chebyshev nodes on the $\kt$-space, whereas $\zeta\left({\kt}_i\right)$ is the projection of the momentum $\kt$ on the interval of $[-1,1]$ interval, suitable for Chebyshev interpolation.  The integers $N_x$ and $N_k$ are the numbers of grid points in $x$ and $\kt$ ranges respectively. Finally we define the function  $t\left(x,n\right)$ which is a combination of  triangular functions providing the  trapezoidal interpolation in $\ln{x}$:
		\begin{equation}
		\begin{split}
		t\left(x,n\right)=&\ln{\left(x/x_{n+1}\right)}/\ln{\left(x_n/x_{n+1}\right)}\Theta\left(x-x_{n+1}\right)\Theta\left(x_n-x\right)\\+&\ln{\left(x_{n-1}/x\right)}/\ln{\left(x_{n-1}/x_n\right)}\Theta\left(x_{n-1}-x\right)\Theta\left(x-x_n\right)\;.
		\end{split}
		\end{equation}
		
		Further rewriting~\eqref{eq:num1ststep} into the matrix form results into this form of the equation
		
		We define a matrix
		\begin{equation}
		\begin{split}
		{\bar K}_{ij;mn}=&{\bar\alpha}_s\int\frac{d^2{\bf q}}{\pi{{\bf q}^2}}\left\{t\left(x\frac{{k^2_T}_i+q^2_T}{{k^2_T}_i},n\right)c_m\left(\zeta\left(|{\bf k}_i-{\bf q}|\right),N_k\right)\right.\\ &\qquad -\,\left.\theta\left({k_T}^2_i-q_T^2\right)t\left(x_j,n\right)c_m\left(\zeta\left({k_{T}}_i\right),N_k\right)\right\}\;
		\end{split}
		\end{equation}
		
		and finally write:
		\begin{equation}\label{eq:numderivative}
		\begin{split}
		\frac{\partial{\cal F}(x,{k_T}_i^2)}{\partial\ln{x}}\bigg|_{x=x_j}&=\frac{\partial{\cal F}_0(x,{k_T}_i^2)}{\partial\ln{x}}\bigg|_{x=x_j}-\sum\limits_{m=0}^{N_k-1}\sum\limits_{n=0}^{N_x-1}\,{\bar K}_{ij;mn}\,\mathcal{F}_{mn}\;.
		\end{split}
		\end{equation}
		
		To solve the equation we can use the discretized form of the differential of the function ${\cal F}(x,{k_T}_i^2)$ in $\ln{x}$:
		\begin{equation}\label{eq:differential}
		\begin{split}
		{\cal F}\left(x\,(1+d\ln{x}),{k_T}_i^2\right)=\,&{\cal F}\left(x,{k_T}_i^2\right)+\frac{\partial{\cal F}\left(x,{k_T}_i^2\right)}{\partial\ln{x}}d\ln{x}\;\\ \downarrow& \\
		{\cal F}\left(x_{j+1},{k_T}_i^2\right)=\,&{\cal F}\left(x_j,{k_T}_i^2\right)+\frac{\partial{\cal F}\left(x,{k_T}_i^2\right)}{\partial\ln{x}}\bigg|_{x=x_j}\ln{\left(\frac{x_{j+1}}{x_j}\right)}\;.
		\end{split}
		\end{equation}
		
		The equation above~\eqref{eq:differential} together with~\eqref{eq:numderivative} with the boundary condition ${\cal F}\left(x_0,{k_T}_i^2\right)={\cal F}_0\left(x_0,{k_T}_i^2\right)$ can be used to solve the equation~\eqref{eq:unresmBFKLkcdiffR} numerically using the iteration method. The algorithm can be summarized in the following diagram:
		\begin{equation}\label{eq:numdiagram}
		\begin{split}
		&\qquad{\cal F}\left(x_j,{k_T}_i^2\right)\\
		&\eqref{eq:numderivative}\downarrow\qquad\qquad\uparrow\eqref{eq:differential}\\
		&\qquad\frac{\partial{\cal F}(x,{k_T}_i^2)}{\partial\ln{x}}\bigg|_{x=x_j}
		\end{split}
		\end{equation}
		
		Distance between the values of ${\cal F}\left(x_j,{k_T}_i^2\right)$ in two subsequent cycles is evaluated after each cycle and the algorithm is terminated when the distance becomes sufficiently small.
		
		We parametrize the grid points in following ways:
		\begin{equation}
		\begin{split}
		x_j=\,&x_{max}\exp{\left[-\frac{j}{N_x-1}\log\frac{x_{max}}{x_{min}}\right]}\;,\\
		{k_T}_i=\,&Q_0\,\exp\left[\frac{\zeta_i+1}{2}\log\frac{w}{Q_0}\right]\;,
		\end{split}
		\end{equation}
		
		with $x_{max}$ and $x_{min}$ being the upper and lower limit of the $x$-axis of the grid, $w$ and $Q_0$ being the upper and lower limit of the $k_T$-axis of the grid and $\zeta_i=\cos\left(\frac{\pi(i+1/2)}{N_k}\right)$ being the Chebyshev node.

\end{document}